\pdfoutput=1
\documentclass[aps,superscriptaddress,twocolumn,prd,nofootinbib,longbibliography,floatfix]{revtex4-2}
\usepackage{amsmath, amssymb, amsfonts, amsthm, latexsym, epsfig, mathrsfs, xcolor, bbm, slashed,braket,bm}

\usepackage[inline]{enumitem}

\usepackage{setspace}
\usepackage[marginal, multiple]{footmisc}

\usepackage[T1]{fontenc}
\usepackage[utf8]{inputenc}
\usepackage{lmodern}
\usepackage[normalem]{ulem}

\usepackage[unicode, colorlinks, allcolors=blue!70!black, linktocpage, pdfusetitle]{hyperref}
\usepackage[all]{hypcap}

\definecolor{indigo(dye)}{rgb}{0.0, 0.25, 0.42}

\usepackage{tikz}
\usetikzlibrary{arrows.meta}

\numberwithin{equation}{section}

\usepackage{cleveref}

\usepackage{cancel}

\usepackage{microtype}

\setlist[enumerate]{noitemsep, label=(\arabic*), ref=(\arabic*)}

\newlist{condlist}{enumerate}{2}
\setlist[condlist,1]{noitemsep, label=(\arabic*), ref=(\arabic*)}
\setlist[condlist,2]{noitemsep, label=(\alph*), ref=(\arabic{condlisti}.\alph*)}
\crefname{condlisti}{condition}{conditions}
\crefname{condlistii}{condition}{conditions}

\renewcommand\thesection{\arabic{section}}
\renewcommand\thesubsection{\arabic{subsection}}

\makeatletter
\def\p@subsection{\thesection.}
\def\p@subsubsection{\thesection.\thesubsection.}
\makeatother

\theoremstyle{plain}

\theoremstyle{definition}

\theoremstyle{remark}

\crefname{equation}{Eq.}{Eqs.}
\Crefname{equation}{Equation}{Equations}

\crefname{section}{Sec.}{Secs.}
\crefname{appendix}{Appendix}{Appendices}
\crefname{figure}{Fig.}{Figs.}

\crefname{definition}{Def.}{Defs.}
\crefname{prop}{Prop.}{Props.}
\crefname{lemma}{Lemma}{Lemmas}
\crefname{corollary}{Cor.}{Cors.}
\crefname{thm}{Theorem}{Theorems}
\crefname{remark}{Remark}{Remarks}

\crefname{ass}{Assumptions}{Assumptions}
\crefname{property}{Properties}{Properties}

\newcommand{\be}{\begin{equation}}
\newcommand{\ee}{\end{equation}}

\newcommand{\mc}{\mathcal}

\newcommand{\ms}{\mathscr}

\newcommand{\bb}{\mathbb}

\newcommand{\Lie}{\pounds}
\newcommand{\hatLie}{\Lie\kern-0.25em\hat{\vphantom{\Lie{}}}\kern0.25em}

\let\oldint\int
\renewcommand{\int}{\oldint\limits}

\newcommand{\op}[1]{\boldsymbol{#1}}

\newcommand{\scri}{\ms I}

\begin{document}

\title{Killing Horizons Decohere Quantum Superpositions}
\author{Daine L. Danielson}\email{daine@uchicago.edu}
\affiliation{Enrico Fermi Institute and Department of Physics, The University of Chicago, 933 East 56th Street, Chicago, Illinois 60637, USA}
\author{Gautam Satishchandran}\email{gautam.satish@princeton.edu}
\affiliation{Princeton Gravity Initiative, Princeton University, Jadwin Hall, Washington Road, Princeton NJ 08544, USA}
\affiliation{Enrico Fermi Institute and Department of Physics, The University of Chicago, 933 East 56th Street, Chicago, Illinois 60637, USA}
\author{Robert M. Wald}\email{rmwa@uchicago.edu}
\affiliation{Enrico Fermi Institute and Department of Physics, The University of Chicago, 933 East 56th Street, Chicago, Illinois 60637, USA}

\date{\today}

\begin{abstract}

\noindent We recently showed that if a massive (or charged) body is put in a quantum spatial superposition, the mere presence of a black hole in its vicinity will eventually decohere the superposition. In this paper we show that, more generally, decoherence of stationary superpositions will occur in any spacetime with a Killing horizon. This occurs because, in effect, the long-range field of the body is registered on the Killing horizon which, we show, necessitates a  flux of ``soft horizon gravitons/photons'' through the horizon. The Killing horizon thereby harvests ``which path'' information of quantum superpositions and will decohere any quantum superposition in a finite time. It is particularly instructive to analyze the case of a uniformly accelerating body in a quantum superposition in flat spacetime. As we show, from the Rindler perspective the superposition is decohered by ``soft gravitons/photons'' that propagate through the Rindler horizon with negligible (Rindler) energy. We show that this decoherence effect is distinct from---and larger than---the decoherence resulting from the presence of Unruh radiation.
We further show that from the inertial perspective, the decoherence is due to the radiation of high frequency (inertial) gravitons/photons to null infinity. (The notion of gravitons/photons that propagate through the Rindler horizon is the same notion as that of gravitons/photons that propagate to null infinity.) We also analyze the decoherence of a spatial superposition due to the presence of a cosmological horizon in de Sitter spacetime.
We provide estimates of the decoherence time for such quantum superpositions in both the Rindler and cosmological cases. Although we explicitly treat the case of spacetime dimension $d=4$, our analysis applies to any dimension $d \geq 4$.
\end{abstract}
\maketitle

\section{Introduction}

Consider a stationary spacetime in which an experimentalist, Alice, is present. Alice's lab is stationary, and she has control of a charged or massive body (hereinafter referred to as a ``particle''). She sends her particle through a Stern-Gerlach apparatus or other device that puts her particle in a quantum superposition of two spatially separated states\footnote{Quantum spatial superpositions of massive bodies have been of recent interest in both theoretical as well as proposed experimental probes of fundamental properties of quantum gravity, e.g., \cite{Bose_2017,Marletto_2017,Belenchia_2018,Christodoulou_2019,Giacomini_2019,Aspelmeyer_2021,Danielson_2021,Carney_2021,christodoulou_2022,carney_2022,Feng_2022,Zhou_2022,Overstreet_2022}.}. She keeps these spatially separated components stationary for a time $T$ and then recombines them. Will Alice be able to maintain the coherence of these components, so that, when recombined, the final state of her particle will be pure---or will decoherence have occurred, so that the final state of her particle will be mixed?

Ordinarily, any decoherence effects will be dominated by ``environmental influences,'' i.e., additional degrees of freedom present in Alice's lab that interact with her particle. We assume that Alice has perfect control of her laboratory and its environment so that there is no decoherence from ordinary environmental effects. However, for a charged or massive particle, Alice cannot perfectly control the electromagnetic or gravitational field, since her particle acts as a source for these fields and some radiation will be emitted during the portions of her experiment where she separates and recombines her particle. Nevertheless, in Minkowski spacetime, if her lab is stationary in the ordinary, inertial sense, she can perform her experiment in a sufficiently adiabatic manner that negligible decohering radiation is 
emitted. In principle, she can keep the particle separated for an arbitrarily long time $T$ and still maintain coherence when the components are recombined.

In a recent paper \cite{Danielson_2022}, we showed that the above situation changes dramatically if a black hole is present in the spacetime---even though the experiment is carried out entirely in the black hole's exterior. In effect, a black hole horizon harvests ``which path'' information about any quantum superposition in its exterior, via the long-range fields sourced by the superposed matter. We showed that this results in the unavoidable radiation of entangling ``soft photons or gravitons'' through the horizon that carry the ``which path'' information into the black hole. Consequently, the mere presence of the black hole implies a fundamental rate of decoherence on the quantum superposition\footnote{In QED, this effect applies only to superpositions of charged particles. However, since all matter sources gravity, the quantum gravitational decoherence applies to all superpositions.}.  Although the rate of decoherence will be small if the black hole is far away, the coherence decays exponentially in the time, $T$, that the spatial superposition is maintained. Thus, in any spacetime with a black hole, there will be essentially complete decoherence within a finite time\footnote{This maximal coherence time for superpositions in the exterior can be much smaller than the evaporation time of the black hole.}.

The purpose of this paper is to generalize the results of \cite{Danielson_2022} to spacetimes with Killing horizons, i.e., spacetimes with a Killing vector field such that there is a null surface to which the Killing field is normal (see, e.g., \cite{Kay:1988mu} for a discussion of properties of Killing horizons). The event horizon of a stationary black hole is a Killing horizon \cite{HawkingEllis:1973,Hawking:1971vc,Alexakis:2009gi}, so spacetimes with Killing horizons encompass the case of stationary spacetimes that contain black holes. However, there are many cases of interest where Killing horizons are present without the presence of black holes. One such case is that of Minkowski spacetime, where the Rindler horizon is a Killing horizon with respect to the Lorentz boost Killing field. Another such case is de Sitter spacetime, where the cosmological horizon is a Killing horizon. We will show that in these cases, a spatial superposition that is kept stationary (with respect to the symmetry generating the Killing horizon) will decohere in a manner similar to the black hole case. We will also provide an estimate of the maximum amount of time during which coherence can be maintained.

The case of the Rindler horizon is particularly instructive. The relevant symmetry here is that of Lorentz boosts, so Alice's lab will be ``stationary'' if it is following orbits of Lorentz boosts, which are uniformly accelerating worldlines. Our analysis based upon radiation through the Rindler horizon shows that decoherence of a uniformly accelerating spatially separated superposition occurs because of 
the emission of ``soft'' (i.e., very low frequency) gravitons or photons, where the frequency is defined relative to an affine parameter on the Rindler horizon. As we shall show, the decoherence effect of this radiation of soft gravitons or photons is distinct from the (smaller) decoherence effect resulting from the presence of Unruh radiation. To gain further insight, we also
analyze the decohering radiation in the electromagnetic case from the inertial point of view, using the Li\'enard-Wiechert solution to determine the radiation at future null infinity. As we shall show, the decohering photons are of high frequency at null infinity. 

In sec.~\ref{raddecoh} we provide a general discussion of the decoherence of a quantum superposition due to radiation in a stationary spacetime. In sec.~\ref{subsec:decacc} we consider the decoherence of a uniformly accelerating superposition, analyzing it from both the Rindler and Minkowski viewpoints. We also show that this decoherence is distinct from (and larger than) the decoherence effects due to the presence of Unruh radiation. In sec.~\ref{cosdecoh} we analyze the decoherence in de Sitter spacetime associated with the cosmological horizon. We will work in Planck units where $G=c=\hbar=k_{\textrm{B}}=1$ and, in electromagnetic formulas, we also put $\epsilon_0 = 1$, but we will restore these constants in our formulas that give estimates for decoherence times. Lower case Latin indices represent abstract spacetime indices. Upper case Latin indices from the early alphabet correspond to spatial indices on horizons or null infinity.

\section{Decoherence Due to Radiation in a Stationary Spacetime}
\label{raddecoh}

In this section, we will give a general analysis of the decoherence of a spatial superposition in a stationary spacetime due to emission of radiation by the body. Our analysis applies both to the decoherence of a charged body due to emission of electromagnetic radiation and to the decoherence of a gravitating body due to emission of linearized gravitational radiation. The analyses of these two cases are very closely parallel. In order to avoid repetition, we will analyze only the electromagnetic case in detail, but near the end of this section, we will state the corresponding results in the linearized gravitational case, which can be obtained straightforwardly by replacing the vector potential $A_a$ with the perturbed metric $h_{ab}$, the charge-current $j_a$ with the stress-energy $T_{ab}$, etc.

Consider a charged particle\footnote{As already indicated above, the ``particle'' need not be an elementary particle but could be a ``nanoparticle'' or any other body whose only relevant degree of freedom for our analysis is its center of mass.} in a stationary spacetime. We assume that the particle is initially in a stationary state. The particle is then put through a Stern-Gerlach (or other) apparatus, resulting in it being in a superposition state\footnote{For simplicity, we have assumed that we have a $50$-$50$ superposition of $\ket{\psi_1}$ and $\ket{\psi_2}$, but this assumption is not necessary.}
\be
\ket{\psi} = \frac{1}{\sqrt{2}} \left(\ket{\psi_1} + \ket{\psi_2} \right)
\label{sup12}
\ee
where $\ket{\psi_1}$ and $\ket{\psi_2}$ are normalized states that are spatially separated after passing through the apparatus. The particle is then recombined via a reversing Stern-Gerlach (or other) apparatus and returns to a stationary state. We are particularly interested in the case where, between separation and recombination, $\ket{\psi_1}$ and $\ket{\psi_2}$ are kept stationary for a long period of time, $T$, but we do not make any such assumption in this section. We wish to estimate how much decoherence due to emission of electromagnetic radiation will have occurred by the time of recombination\footnote{\label{decft}The decoherence of Alice's particle can be experimentally determined as follows. We assume that Alice's particle initially has spin in the positive $x$-direction and thus is in a $50$-$50$ superposition of $z$-spin after passing through the initial Stern-Gerlach apparatus. After recombination, Alice measures the $x$-spin of her particle. If coherence of the superposition eq.~(\ref{sup12}) has been maintained, then (assuming that Alice has made appropriate corrections if there are any phase differences between the paths) the spin will always be found to be in the positive $x$-direction. On the other hand, if any
coherence has been lost, the particle will not be in a state of definite spin, and the spin will sometimes be found to be in the negative $x$-direction. By repeating the experiment many times, Alice can, in principle, determine the decoherence to any desired accuracy.}.

A key assumption that we shall make is that the fluctuations in the charge-current operator $\boldsymbol{j}^a$ in the states $\ket{\psi_1}$ and $\ket{\psi_2}$ are negligibly small over the scales of interest so that we can treat the charge current in each of these states as $c$-number sources in Maxwell's equations, given by $j_1^a = \langle \psi_1| \boldsymbol{j}^a | \psi_1 \rangle$ and $j_2^a = \langle \psi_2| \boldsymbol{j}^a | \psi_2 \rangle$, respectively. In the initial and final stationary eras, $\ket{\psi_1}$ and $\ket{\psi_2}$ are assumed to coincide spatially (though they may differ in other characteristics, such as spin) so that $j_1^a = j_2^a$ at very early and very late times. 

In order to proceed further, we must specify the initial state of the electromagnetic field. Since, prior to going through the Stern-Gerlach apparatus, the charge is assumed to be stationary, at early times we may subtract the ``Coulomb field'' $C^{\rm in}_a$ of the charge, i.e., at early times we may consider the electromagnetic field observable 
\be
\boldsymbol{A}^{\rm in}_a = \boldsymbol{A}_a - C^{\rm in}_a  \boldsymbol{1}
\ee
where $C^{\rm in}_a$ is the (assumed to be unique) stationary classical solution to Maxwell's equations with the early time stationary charged particle source $j_1^a = j_2^a$ and $\bm{A}_{a}$ is the vector potential operator. We need not assume any specific choice of gauge for $\boldsymbol{A}^{\rm in}_a$. Then $\boldsymbol{A}^{\rm in}_a$ satisfies the source-free Maxwell's equations at early times, and we may extend its definition to all times by requiring it to satisfy the source-free Maxwell equations everywhere. 

The initial state of the electromagnetic field may be specified by 
giving the ``radiation state'' of $\boldsymbol{A}^{\rm in}_a$. The choice of this state depends on the physical situation being considered.
If the spacetime were globally stationary---i.e., if the stationary Killing field were everywhere timelike, so, in particular, there are no Killing horizons---it would be natural to assume that the initial state of the radiation is the stationary vacuum state, i.e., the ground state relative to the time translations. For the case of a black hole spacetime, it would be correspondingly natural to assume that the initial state of the radiation is that of the Unruh vacuum, since for a black hole formed by gravitational collapse, the state of a quantum field is expected to approach the Unruh vacuum after the black hole has ``settled down'' to a stationary state. For the case of Minkowski spacetime, we take the initial state of the radiation to be the ordinary (inertial) Minkowski vacuum. For de Sitter spacetime, we take the initial state of the radiation to be the de Sitter invariant vacuum\footnote{\label{dsinv} A de Sitter invariant vacuum state does not exist for the massless scalar field \cite{Allen_1985_scalar} but such a state does exist for the electromagnetic field \cite{Allen:1985_vector} and linearized gravitational field \cite{Allen:1986_graviton}.} for the electromagnetic field \cite{Allen:1985_vector}. We denote the initial state of the radiation in all of the above cases by $\ket{\Psi_0}$. 

In each of the above cases, $\ket{\Psi_0}$ is a pure, quasi-free (i.e., Gaussian) state. It follows (see, e.g., \cite{Wald_book_QFTCSBHT} or appendix A of \cite{Kay:1988mu}) that we can construct a one-particle Hilbert space ${\mathcal H}_{\rm in}$ and corresponding Fock space ${\mathcal F}({\mathcal H}_{\rm in})$ wherein $\ket{\Psi_0}$ plays the role of the vacuum state and the field operator $\boldsymbol{A}^{\rm in}_a$ is represented on $\mathcal F({\mathcal H}_{\rm in})$ by 
\be
\label{eq:Acreatannih}
\boldsymbol{A}^{\rm in}_a (f^a) = i \boldsymbol{a}(\overline{K\sigma_f}) - i \boldsymbol{a}^\dagger (K \sigma_f). 
\ee
Here $f^a$ a divergence-free\footnote{Restriction of the smearing to divergence-free test functions is necessary and sufficient to eliminate the gauge dependence of $\boldsymbol{A}^{\rm in}_a$ (see, e.g., P.101 of \cite{Wald_book_QFTCSBHT}).} test function,
$\sigma_f$ denotes the advanced minus retarded solution to Maxwell's equations with source $f^a$, and $K: S \to {\mathcal H}_{{\rm in}}$ denotes the map taking the space $S$ of classical solutions to their representatives in the one-particle Hilbert space ${\mathcal H}_{\rm in}$. The commutator of the creation and annihilation operators in eq.~(\ref{eq:Acreatannih}) is given by
\begin{equation}
\label{eq:CCR}
[\op{a}(\overline{K\sigma_f}),\op{a}^\dagger(K\sigma_g)] = \braket{K\sigma_{f}|K\sigma_{g}}\op{1}.
\end{equation}
where $\braket{K\sigma_{f}|K\sigma_{g}}$ is the inner product on $\mathcal{H}_{{\rm in}}$, which is given by a natural generalization of the Klein-Gordon inner product to electromagnetic fields.

\par For the case of a globally stationary spacetime in the stationary vacuum state, $K \sigma_f$ corresponds to taking the positive frequency part of $\sigma_f$ with respect to the time translations generating the stationary symmetry. For the case of a stationary black hole in the Unruh vacuum state, $K \sigma_f$ corresponds to taking the positive frequency part of $\sigma_f$ with respect to affine time on the past horizon and with respect to Killing time at past null infinity. For Minkowski spacetime in the inertial Minkowski vacuum, $K \sigma_f$ corresponds to taking the positive frequency part of $\sigma_f$ with respect to inertial time translations. Equivalently, $K \sigma_f$, in this case, corresponds to the solution obtained by taking the positive frequency part of the restriction of $\sigma_f$ to any null hyperplane $\mathcal N$ (i.e., any Rindler horizon) with respect to an affine parametrization of the null geodesics generating $\mathcal N$. For de Sitter spacetime in the de Sitter invariant vacuum, $K \sigma_f$ corresponds to the solution obtained by taking the positive frequency part of the restriction of $\sigma_f$ to any cosmological horizon with respect to an affine parametrization of the null geodesics generating that horizon.

Under the above assumption that the charge-currents of $\ket{\psi_1}$ and $\ket{\psi_2}$ can be treated as $c$-number sources, the electromagnetic field $\boldsymbol{A}_{i,a}$ in the presence of the charge in state $\ket{\psi_i}$ for $i=1,2$ is given in terms of the source free field $\boldsymbol{A}^{\rm in}_a$ by \cite{Yang:1950vi}
\be
\boldsymbol{A}_{i,a} = \boldsymbol{A}^{\rm in}_a +  G^{\rm ret}_{a} (j_i^b) \boldsymbol{1}
\ee
where $G^{\rm ret}_a (j_i^b)$ denotes the classical retarded solution for source $j_i^b$. 
In particular, since the field $\boldsymbol{A}^{\rm in}_a$ is in state $\ket{\Psi_0}$, the correlation functions of the electromagnetic field 
$\boldsymbol{A}_{i,a}$ for $\ket{\psi_i}$ are given by\footnote{It is understood that each of the $x_k$ variables should be smeared with a divergence-free test vector field $f^a_k$.}
\begin{align}
\langle \boldsymbol{A}_{i, a_1} (x_1) &\dots \boldsymbol{A}_{i, a_n} (x_n) \rangle \nonumber \\
= \langle \Psi_0| &\left[\boldsymbol{A}^{\rm in}_{a_1}(x_1)+ G^{\rm ret}_{a_1} (j_i^b)(x_1) \boldsymbol{1}) \right]  \nonumber \\
& \dots \relax \left[\boldsymbol{A}^{\rm in}_{a_n}(x_n)  + G^{\rm ret}_{a_n} (j_i^b)(x_n) \boldsymbol{1})\right] | \Psi_0 \rangle.
\label{corrfun}
\end{align}

Equation~(\ref{corrfun}) is valid at all times. However, at late times---i.e., to the future of any Cauchy surface $\Sigma$ corresponding to the time at which recombination has occurred---we can again subtract off the common stationary Coulomb field, $C^{\rm out}_a$, of $j_1^a = j_2^a$ to obtain the source-free field\footnote{Note that $\boldsymbol{A}^{\rm in}_a$ did not have a subscript ``$i$'' whereas $\boldsymbol{A}_{i,a}$ and $\boldsymbol{A}^{\rm out}_{i, a}$ do carry such subscripts. This is a consequence of the fact that we are working in the ``in'' representation---i.e., the Heisenberg representation on the Hilbert space $\mathcal F({\mathcal H}_{\rm in})$---so $\boldsymbol{A}^{\rm in}_a$ does not depend on the sources, but the other fields do.} $\boldsymbol{A}^{\rm out}_{i,a}$ that describes the radiation at late times for the states $\ket{\psi_i}$,
\be
\boldsymbol{A}^{\rm out}_{i, a} = \boldsymbol{A}_{i, a} - C^{\rm out}_a  \boldsymbol{1} \, .
\ee
By eq.~(\ref{corrfun}), at late times, the correlation functions of $\boldsymbol{A}^{\rm out}_a$ are given by
\begin{align}
\langle \boldsymbol{A}^{\rm out}_{i, a_1} (x_1) &\dots \boldsymbol{A}^{\rm out}_{i, a_n} (x_n) \rangle \nonumber \\
= \langle \Psi_0| &\left[\boldsymbol{A}^{\rm in}_{a_1}(x_1)  + {\mathcal A}_{i, a_1}(x_1) \boldsymbol{1}) \right]  \nonumber \\
& \dots \relax \left[\boldsymbol{A}^{\rm in}_{a_n}(x_n)  + {\mathcal A}_{i, a_n} (x_n) \boldsymbol{1})\right] | \Psi_0 \rangle
\label{corrfun2}
\end{align}
where
\be
{\mathcal A}_{i, a} = G^{\rm ret}_{a} (j_i^b) - C^{\rm out}_a.
\label{coulsub}
\ee
Note that ${\mathcal A}_{i, a}$ is a classical solution of the source-free Maxwell equations in the late-time region. 

The correlation functions eq.~(\ref{corrfun2}) on any late-time Cauchy surface are precisely those of the coherent state
\be
\ket{\Psi_i} =e^{-\frac{1}{2}\lVert K\mathcal{A}_{i}\rVert^2} \exp \left[ \boldsymbol{a}^\dagger (K {\mathcal A}_i) \right] \ket{\Psi_0},
\ee  
where the norm is that of the one-particle inner product of eq.~(\ref{eq:CCR}). Thus, the coherent state
$\ket{\Psi_1}$ describes the ``out'' radiation state corresponding to charged particle state $\ket{\psi_1}$ and the coherent state $\ket{\Psi_2}$ describes the ``out'' radiation state corresponding to charged particle state $\ket{\psi_2}$. The joint ``out'' state, $\ket{\Upsilon}$, of the particle-radiation system is given by 
\be
\label{eq:Upsilon}
\ket{\Upsilon} = \frac{1}{\sqrt{2}} \left(\ket{\psi_1} \otimes  \ket{\Psi_1}+ \ket{\psi_2} \otimes  \ket{\Psi_2} \right).
\ee
Therefore, the decoherence of $\ket{\psi_1}$ and $\ket{\psi_2}$ due to emission of electromagnetic radiation is given by
\begin{equation}
\label{eq:decSigma}
    \ms{D}=1-|\braket{\Psi_{1}|\Psi_{2}}|.
\end{equation}
We wish to evaluate $\ms D$.

By the general formula for the inner product of coherent states, we have
\be
|\braket{\Psi_1|\Psi_2}| = \exp \left[- \frac{1}{2} \lVert K({\mathcal A}_{1} - {\mathcal A}_{2 })\rVert^2 \right].
\ee
Now, in the late-time era, ${\mathcal A}_{1, a} - {\mathcal A}_{2, a}$ is just the difference between the classical retarded solutions with sources $j_1^a$ and 
$j_2^a$,
\be
{\mathcal A}_{1, a} - {\mathcal A}_{2,a} = G^{\rm ret}_{a} (j_1^b) - G^{\rm ret}_{a} (j_2^b) = G^{\rm ret}_{a} (j_1^b - j_2^b).
\ee
Consider the coherent state associated with $G^{\rm ret}_{a} (j_1^b - j_2^b)$ in the late-time era. We refer to photons in this state as {\em entangling photons}. By the general properties of coherent states, the expected number, $\langle N \rangle$, of entangling photons is given by
\be
\label{eq:entphot}
\langle N \rangle \equiv \lVert K \left[G^{\rm ret} (j_1 - j_2) \right]\rVert^2.
\ee
Thus, we have
\be
|\braket{\Psi_1|\Psi_2}| = \exp \left[- \frac{1}{2} \langle N \rangle \right]
\ee
so
\begin{equation}
\label{eq:decNE}
    \ms{D}=1-|\braket{\Psi_{1}|\Psi_{2}}| = 1 - \exp \left[- \frac{1}{2} \langle N \rangle \right]
\end{equation}
and we see that the necessary and sufficient condition for significant decoherence ($\mathscr{D}\sim 1$) is $\langle N\rangle \gtrsim 1$.

We summarize the results that we have obtained above as follows. Under the assumptions we have made above, in order to calculate the decoherence, $\ms D$, of the particle due to radiation, we carry out the following steps: 

\begin{enumerate}

\item We obtain the expected charge current, $j_1^a$ and $j_2^a$, for the particle in states $\ket{\psi_1}$ and $\ket{\psi_2}$ of the superposition. 

\item We calculate the classical retarded solution, $G^{\rm ret}_{a} (j_1^b - j_2^b)$ for the difference of these charge currents, which is a source-free solution at late times, since $j_1^a = j_2^a$ at late times. 

\item \label{Kmap} We calculate the one-particle state $K G^{\rm ret} (j_1 - j_2)$ corresponding to $G^{\rm ret}_{a} (j_1^b - j_2^b)$ at late times. In the various cases, this corresponds to the following: (i) For a globally stationary spacetime initially in the stationary vacuum state, this one-particle state is the positive frequency part of the solution with respect to the time translations generating the stationary symmetry. (ii) For the case of a stationary black hole initially in the Unruh vacuum, the one-particle state is the positive frequency part of the solution with respect to affine time on the past horizon and with respect to Killing time at past null infinity. (iii) For Minkowski spacetime initially in the Minkowski vacuum, the one-particle state is the positive frequency part of the solution with respect to inertial time or, equivalently, the positive frequency part with respect to affine time on any Rindler horizon. (iv) For de Sitter spacetime initially in the de Sitter invariant vacuum, the one-particle state is the positive frequency part of the solution with respect to affine time on any cosmological horizon.

\item We compute the squared norm, $\lVert K [G^{\rm ret} (j_1 - j_2)]\rVert^2$, of this one-particle state at late times. This quantity is equal to the expected number of entangling photons, $\langle N \rangle$. The decoherence due to radiation is then given by
\be
\label{decrad}
\ms{D} = 1 -  \exp \left[- \frac{1}{2} \lVert K \left[G^{\rm ret} (j_1 - j_2) \right]\rVert^2 \right].
\ee

\end{enumerate}

As previously stated, the above analysis extends straightforwardly to the linearized gravitational case, where the perturbed metric, $h_{ab}$, is treated as a linear quantum field propagating in the background classical stationary spacetime. To compute the decoherence due to gravitational radiation in this case, we carry out the above steps, replacing $A_a$ by $h_{ab}$ and the charge-current $j^a$ by the stress-energy tensor $T_{ab}$. The retarded solution $G^{\rm ret}_{a} (j^b)$ for Maxwell's equations is replaced by the retarded solution $G^{\rm ret}_{ab} (T_{cd})$ for the linearized Einstein equation. The map $K: S \to {\mathcal H}_{\textrm{in}}$ is again obtained as in \cref{Kmap} above and the inner product on $\mathcal H_{\textrm{in}}$ is again given by a natural generalization of the Klein-Gordon inner product to linearized gravitational fields. The decoherence due to gravitational radiation is then given by the analog of eq.~(\ref{decrad}).

The above analysis applies for any motion of the components of Alice's superposition. We are primarily interested in the case where, during a time interval $T_1$, Alice puts a particle of charge $q$ (or mass $m$) into a spatial superposition, where the distance between the components of the particle wavefunction is $d$. She then keeps this superposition stationary in her lab for a time $T$. Finally, she recombines her particle over a time interval $T_2$. 

In Minkowski spacetime in the case where Alice's lab is inertial, $G^{\rm ret}_{a} (j_1^b - j_2^b)$ will be nonzero at null infinity only at the retarded times corresponding to the time intervals $T_1$ and $T_2$. A rough estimate of the number of entangling photons was obtained in \cite{Belenchia_2018} using the Larmor formula for radiation in these eras, which, in natural units, yields
\begin{equation}
    \braket{N} \sim \frac{q^{2}d^{2}}{  \, [{\rm min}(T_1, T_2)]^{2}}  \quad \textrm{(Minkowski, EM).}
\end{equation}
The corresponding result in the linearized gravitational case is \cite{Belenchia_2018}
\begin{equation}
    \braket{N} \sim \frac{m^{2}d^{4}}{ \, [{\rm min}(T_1, T_2)]^{4}}  \quad \textrm{(Minkowski, GR).}
\end{equation}
Therefore, if Alice recombines her particle sufficiently slowly that $T_1, T_2 \gg qd$ in the electromagnetic case or $T_1, T_2 \gg m d^2$ in the gravitational case, then she can maintain the quantum coherence of her particle. In particular, Alice can keep the components of her particle separated for as long a time $T$ as she likes without destruction of the coherence.

As shown in \cite{Danielson_2022}, the situation is quite different if a black hole is present. In the electromagnetic case, even if $T_1, T_2 \gg qd$ so that a negligible number of entangling photons is emitted to infinity, there will be entangling radiation emitted into the black hole. For large $T$, the number of entangling photons increases with $T$ as\footnote{In the analysis of \cite{Danielson_2022}, we used the fact that the Unruh vacuum is well approximated by the Hartle-Hawking vacuum at low frequencies near the horizon of the black hole.}
\begin{equation}
\label{eq:NEBh}
    \braket{N}\sim \frac{M^{3}q^{2}d^{2}}{D^{6}}T  \quad \quad \textrm{(black hole, EM)}
\end{equation}
 where $M$ is the mass of the black hole, $D$ is the proper distance of Alice's lab from the horizon of the black hole, and we assume that $D \gtrsim M$. 
The corresponding result in the linearized gravitational case is
 \begin{equation}
    \braket{N}\sim \frac{ M^5 m^2 d^4}{ D^{10}}T  \quad \quad \textrm{(black hole, GR).}
    \label{NGRbh}
\end{equation}
 Thus, the coherence of Alice's particle will always be destroyed within a finite time.

In the next two sections, we will apply the above analysis to the cases of Rindler spacetime and de Sitter spacetime. Although we will explicitly analyze only the Rindler and de Sitter cases, it will be clear from our analysis of the next two sections---as well as our analysis in \cite{Danielson_2022}---that it can be applied to any Killing horizon, provided only that the initial ``vacuum state'' $\ket{\Psi_0}$ of the electromagnetic and/or linearized gravitational field corresponds to one-particle states that are positive frequency with respect to affine time on the future Killing horizon.

\section{Rindler Horizons Decohere Quantum Superpositions}
\label{subsec:decacc}

We now consider the case of Minkowski spacetime\footnote{We explicitly treat the case of $4$ spacetime dimensions, but our analysis generalizes straightforwardly to all higher dimensions.} with Alice's lab uniformly accelerating with acceleration $a$. Specifically, we take Alice's lab to follow the orbit 
\be
t = \frac{1}{a}\sinh(a\tau), \quad \quad z = \frac{1}{a}\cosh(a\tau)
\ee
of the boost Killing field 
\begin{equation}
    b^{a} = a\bigg[z\bigg(\frac{\partial}{\partial t}\bigg)^{a} + t \bigg(\frac{\partial}{\partial z}\bigg)^{a}\bigg].
    \label{boostkvf}
\end{equation}
Here we have normalized $b^{a}$ such that $b^{a}b_{a}=-1$ on the worldline of Alice's laboratory. Thus, $b^{a}$ is the four-velocity of Alice's laboratory and $\tau$ is the proper time in her lab. We introduce the null coordinates
\begin{equation}
\label{eq:UV}
    U\equiv t-z, \quad \quad  V\equiv t+z
\end{equation}
and the corresponding vector fields 
\begin{equation}
    n^{a}\equiv (\partial /\partial V)^{a}, \quad \quad \ell^{a}\equiv (\partial /\partial U)^{a},
\end{equation}
which are globally defined, future-directed null vector fields that satisfy $\ell^{a}n_{a}=-1$. In terms of these coordinates, the Minkowski spacetime metric is
\be
 \eta = -dUdV + dx^2 + dy^2
\ee
and the boost vector field is given by 
\begin{equation}
    b^{a}= a\big[-U\ell^{a} + V n^{a}\big].
\end{equation}
The boost Killing field is null on the two ``Rindler horizons,'' i.e., the two null planes $U=0$ and $V=0$, which divide Minkowski spacetime into four wedges. The orbits of the boost Killing field are future-directed and timelike within the ``right Rindler wedge'' $\mc{W}_{\text{R}}$ which is the region $U<0$ and $V>0$. Thus, the ``right Rindler wedge'' $\mc{W}_{\text{R}}$---where Alice performs her experiment---is a static, globally hyperbolic spacetime where the notion of ``time translations'' is defined by Lorentz boosts. 

We refer to the null surface $U=0$ as the future Rindler horizon and denote it as $\ms{H}_{\text{R}}^{+}$. On the region $V > 0$ of $\ms{H}_{\text{R}}^{+}$, it is useful to introduce the coordinate $v$ by
\begin{equation}
\label{eq:UuVv}
 V=V_{0}e^{av} \quad 
\end{equation}
where $V_{0}$ is an arbitrary constant. Then, for $V > 0$ on $\ms{H}_{\text{R}}^{+}$, we have
\begin{equation}
    \label{eq:bana}
b^{a} \big\vert_{\ms{H}_{\text{R}^{+}}} = aV \bigg(\frac{\partial}{\partial V}\bigg)^{a}\bigg\vert_{\ms{H}_{\text{R}^{+}}} = \bigg(\frac{\partial}{\partial v}\bigg)^{a}\bigg\vert_{\ms{H}_{\text{R}^{+}}} \, .
\end{equation}
Since $(\partial/\partial V)^a$ on the horizon is tangent
to the affinely parameterized null geodesic generators of $\ms{H}_{\text{R}}^{+}$, we refer to $V$ as the ``affine time'' on $\ms{H}_{\text{R}}^{+}$, whereas we refer to $v$ as the ``boost Killing time'' on $\ms{H}_{\text{R}}^{+}$.

\subsection{Decoherence Due to Radiation of Soft Photons/Gravitons Through the Rindler Horizon}
\label{subsec:softrad}

We are now in position to apply the results of sec.~\ref{raddecoh} to the Rindler case. We will first analyze the electromagnetic case and then give the corresponding results in the gravitational case. 

We assume that the electromagnetic field is initially in the Minkowski vacuum state.
We assume that Alice possesses a charged particle that is initially stationary (with respect to the boost Killing field) in her (uniformly accelerating) lab. She then creates a quantum spatial superposition which is held stationary (with respect to the boost Killing field) for a proper time $T$ and is then recombined. We wish to know the degree of decoherence of Alice's particle due to emission of radiation. We may directly apply the analysis of sec.~\ref{raddecoh} to answer this question. 

The future Rindler horizon $\ms{H}_{\text{R}}^{+}$ ($U=0$) does not meet the technical requirements of being a Cauchy surface for Minkowski spacetime, since there are inextendible timelike curves that remain in the past of $\ms{H}_{\text{R}}^{+}$ as well as inextendible timelike curves that lie in the future of $\ms{H}_{\text{R}}^{+}$. However, as argued in \cite{Unruh_1984}, it is effectively a Cauchy surface for determining evolution of solutions to the wave equation. This is most easily seen in the conformally completed spacetime, where $\ms{H}_{\text{R}}^{+}$ is the past light cone of a point $p \in \scri^+$ except for the single generator that lies on $\scri^+$ and it also is the future light cone of a point on $p' \in \scri^-$ except for the single generator that lies on $\scri^-$. Data on the full past light cone of $p$ would determine a solution to the past of $\ms{H}_{\text{R}}^{+}$ and data on the full future light cone of $p'$ would determine a solution to the future of $\ms{H}_{\text{R}}^{+}$, thereby determining a solution everywhere in Minkowski spacetime. However, for solutions with appropriate decay, the data on the missing null geodesic generators of $\scri^+$ and $\scri^-$ can be determined by continuity from the data on $\ms{H}_{\text{R}}^{+}$. Consequently, data on $\ms{H}_{\text{R}}^{+}$ suffices to uniquely characterize solutions with appropriate decay. 
Consequently, the ``out'' states $\ket{\Psi_1}$ and $\ket{\Psi_2}$ of the radiation are completely determined by data on $\ms{H}_{\text{R}}^{+}$. Note that this contrasts sharply with the black hole case, where one would need data on both the future event horizon and future null infinity to characterize the ``out'' state of radiation.

\begin{figure}
    \includegraphics[width=\linewidth]{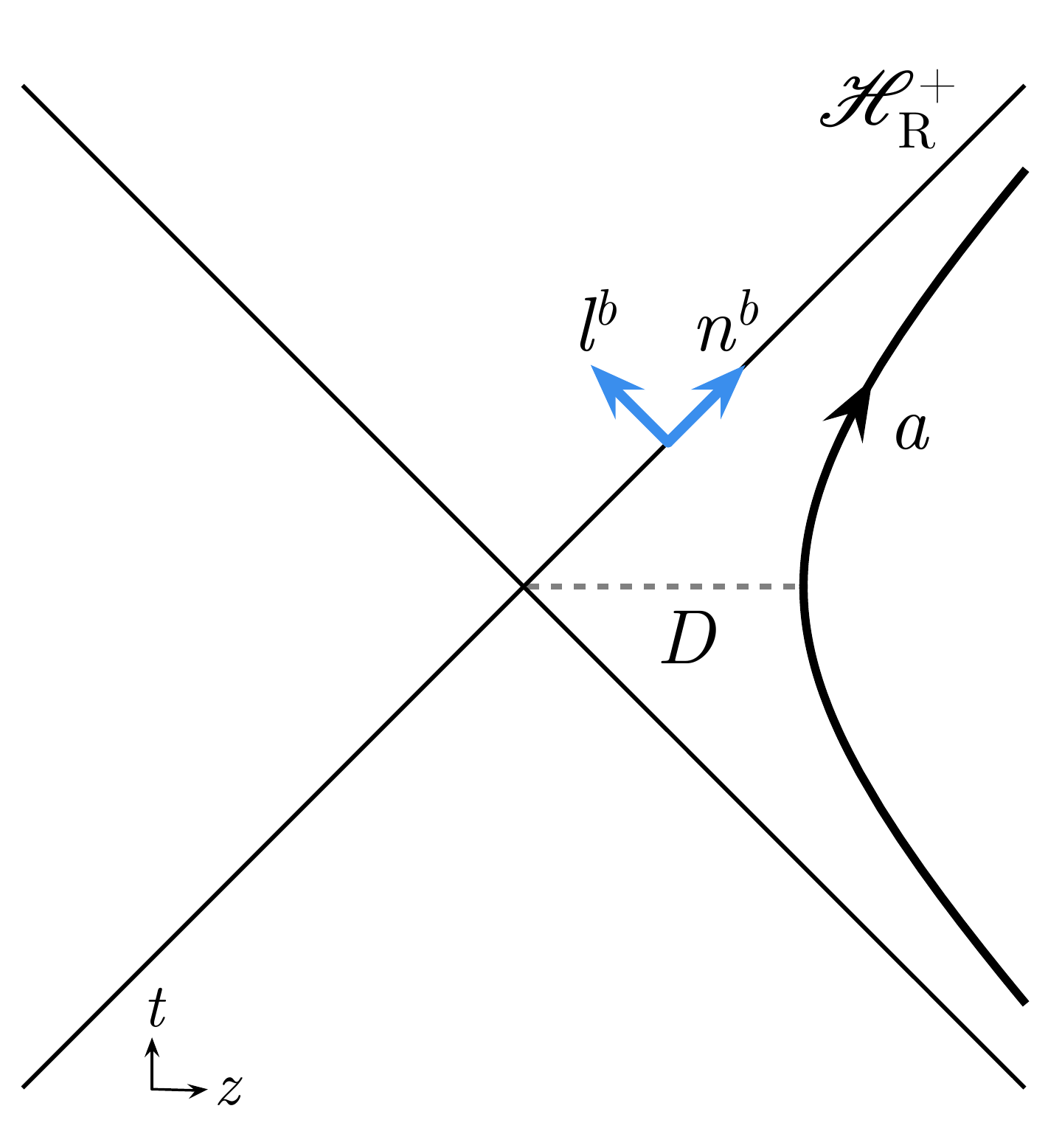}
     \caption{Alice's laboratory undergoes uniform acceleration $a$ in the $z$-direction in Minkowski spacetime and thus follows an orbit of a boost Killing field. The future Rindler horizon $\mathscr H^+_\text{R}$ is a Killing horizon for this boost Killing field. The future-directed null vector $n^b=(\partial/\partial V)^b$ points along the horizon, while $l^b=(\partial/\partial U)^b$ is transverse to it. $D$ is the proper distance from Alice's lab to the horizon. \label{fig:rindler_horizon}}
\end{figure}

The decoherence of Alice's particle due to radiation is given by eq.~(\ref{eq:decNE}). In order to evaluate this, we
first consider a classical point charge of charge $q$ in the ``right Rindler wedge'' $\mc{W}_{\text{R}}$ that is stationary with respect to the boost Killing field and lies at proper distance $D$ from the bifurcation surface of the Rindler horizon. Such a charge will be uniformly accelerating with acceleration $a$ given by
\be
a = \frac{1}{D} \, ,
\ee
as depicted in \cref{fig:rindler_horizon}.

The explicit solution for such a stationary charge in the Rindler wedge has long been known \cite{Whittaker_1927,Bondi_1955,Rohrlich_1961,Boulware_1980,Padmanabhan:2009,ERIKSEN_2004}. The only nonvanishing component of the electromagnetic field in the region $V > 0$ of $\ms{H}_{\text{R}}^{+}$ is 
\begin{equation}
\label{eq:EUptcharge}
    E_{U} \equiv F_{ab}\ell^{a}n^{b}=\frac{2a^{2}q}{ \pi (1+a^{2}\rho^{2})^{2}}
\end{equation}
where $\rho^{2}\equiv x^{2}+y^{2}$. Electromagnetic radiation through the Rindler horizon is described by the pullback, $E_{A}$, of the electric field\footnote{The electric field as measured by an observer with $4$-velocity $u^b$ is $F_{ab} u^b$. Although $n^b$ is null rather than timelike, it is natural (and standard) to use the terminology ``electric field'' for $F_{ab} n^b$ on the horizon.} $E_{a}=F_{ab}n^{b}$ to $\ms{H}_{\text{R}}^{+}$, where the capital Latin indices from the early alphabet denote spatial components in the $x$ and $y$ directions. Since $E_{A}=0$ on the horizon for a uniformly accelerated charge, one may say that a charge held stationary in Alice's lab does not produce any radiation as determined on $\ms{H}_{\text{R}}^{+}$---even though a uniformly accelerated charge radiates (inertial) energy to future null infinity\footnote{A uniformly accelerating charge has a nonvanishing inertial energy current flux $T_{ab} t^a$ through both $\ms{H}_{\text{R}}^{+}$ and $\scri^+$, where $t^a$ denotes a Minkowski time translation. However, the flux of ``boost energy'' $T_{ab} b^a$ vanishes at both $\ms{H}_{\text{R}}^{+}$ and $\scri^+$.}. 

Now consider the case where the point charge is initially uniformly accelerating with acceleration $a$ at a proper distance $D = 1/a$ from the bifurcation surface of the Rindler horizon. The charge is then moved in the $z$-direction\footnote{We consider a $z$-displacement for simplicity. Similar results would hold if the charge were displaced in the $x$ or $y$ directions.} to a different orbit of the same boost Killing field, so that it has uniform acceleration $a'$ and lies at proper distance $D' = 1/a'$ from the Rindler horizon. After the charge has reached its new location, the electric field on $\ms{H}_{\text{R}}^{+}$ is again given by eq.~(\ref{eq:EUptcharge}), but its value, $E'_{U}$, will be different from its value at early times. Maxwell's equations on $\ms{H}_{\text{R}}^{+}$ imply that
\begin{equation}
\label{eq:MaxEq}
    \mathcal{D}^{A}E_{A}=\partial_{V}E_{U}
\end{equation}
where $\mathcal{D}_{A}$ is the derivative operator on the $\bb{R}^{2}$ cross-sections of the horizon and capital Latin indices from the early alphabet are raised and lowered with the metric, $\delta_{AB}$, on the cross sections. Eq.~(\ref{eq:MaxEq}) implies that $E_A \neq 0$ whenever $\partial_{V}E_{U} \neq 0$, so there will be radiation through the horizon as the charge is being moved. Most importantly, it implies that
\be
\mathcal{D}^A \left(\int_{-\infty}^\infty dV E_{A} \right) = \Delta E_U
\label{vpchange}
\ee
where $\Delta E_U = E'_U - E_U$ is the change in the radial electric field between the charge at positions $D'$ and $D$. Now, in a gauge where $A_{a}n^{a} = 0$ on the horizon, the transverse (i.e., $x$-$y$) components of the electric field are related to the corresponding components of the vector potential by
\be
E_{A}=-\partial_{V} A_{A}.
\ee
Since the transverse components of the Coulomb field of a static charge vanish, we may replace the vector potential $A_A$ by the ``Coulomb subtracted'' vector potential $\mc{A}_A$ defined by eq.(\ref{coulsub}), so we have
\be
E_{A}=-\partial_{V} \mc{A}_{A}.
\ee
It then follows immediately from eq.~(\ref{vpchange}) that the difference, $\Delta \mc{A}_A$, between the final and initial values of $\mc{A}_A$ is given by
\be
\mathcal{D}^A (\Delta \mc{A}_A)  = -  \Delta E_U
\label{hormem} 
\ee
independently of the manner in which the charge is moved from $D$ to $D'$. Equation (\ref{hormem}) is an exact mathematical analog of the electromagnetic memory effect at null infinity \cite{Bieri_2013}. 
For the explicit solution eq.~(\ref{eq:EUptcharge}), we have 
\begin{equation}
    \Delta E_{U} \approx \frac{-4qda^{3}(1-a^{2}\rho^{2})}{\pi(1+a^{2}\rho^{2})^{3}}.
\end{equation}
where $d = D' - D$ and we have assumed that
\be
d \ll D = \frac{1}{a} \, .
\label{dllD}
\ee 
From eq.~(\ref{hormem}), we find that $\Delta \mc{A}_A$ points in the $\hat{\rho}$-direction and has magnitude
\begin{equation}
    |\Delta \mc{A}_A| = \Delta \mc{A}_{\rho} \sim\frac{ qd a^3 \rho}{(1 + a^2 \rho^2)^2}.
\label{Aivalue}
\end{equation}
The key point is that even though $E_A=0$ at both late and early times, $\mc{A}_{A}$ does not return to its original value at late times, and the change, $\Delta \mc{A}_A$, in the vector potential between late and early times is determined only by the initial and final positions of the charge.

We now consider the quantized radiation through the horizon resulting from the displacement of the charge, assuming that, after the displacement, the charge is held at its new position, $D'$, forever. For the Fock space associated with the Minkowski vacuum state, the map $K: S \to {\mathcal H}_{\textrm{in}}$ that associates one-particle states to classical solutions is given by taking the positive frequency part of the classical solution with respect to inertial time, with the inner product on ${\mathcal H}_{\textrm{in}}$ given by the Klein-Gordon product. For the electromagnetic field on $\ms{H}_{\text{R}}^{+}$ in a gauge where $\mc{A}_{a}n^{a}=0$ on $\ms{H}_{\text{R}}^{+}$, the ``free data'' on $\ms{H}_{\text{R}}^{+}$ is the pull-back, $\mc{A}_{A}$, of the vector potential. For two classical solutions with data $\mc{A}_{1,A}$ and $\mc{A}_{2,A}$ on $\ms{H}_{\text{R}}^{+}$, the inner product of their corresponding one-particle states is given by 
\cite{Kay:1988mu,Dappiaggi:2017kka} 
\begin{equation}
\label{eq:innerprod0}
    \braket{K\mc{A}_{1}| \,K \mc{A}_{2}}_{\ms{H}_{\text{R}}^{+}}=2\int_{\bb{R}^{2}}dxdy \int_{0}^{\infty}\frac{\omega d\omega}{2\pi}\delta^{AB}\overline{{\hat{\mc{A}}_{1,A}}} \hat{\mc{A}}_{2,B}
\end{equation}
where $\hat{\mc{A}}_{A}(\omega,x^{B})$ is the Fourier transform of $\mc{A}_{A}(V,x^{B})$ with respect to the affine parameter $V$. By the same reasoning as led to eq.~(\ref{eq:entphot}), the expected number of photons on ${\ms{H}^{+}_{\text{R}}}$ in the coherent state associated to any classical solution $\mc{A}_{A}$ is simply
\begin{equation}
    \braket{N} = \lVert K \mc{A}\rVert^2_{\ms{H}^{+}_{\text{R}}}
    \label{Nhor}
\end{equation}
where the norm is defined by the inner product eq.~(\ref{eq:innerprod0}). 
However, since $\Delta \mc{A}_{A} \neq 0$, the Fourier transform, $\hat{\mc{A}}_{A}(\omega,x^{B})$, of $\mc{A}_{A}$ diverges as $1/\omega$ as $\omega \to 0$. It follows that the integrand of the expression for the norm given by the right side of eq.~(\ref{eq:innerprod0}) also diverges as $1/\omega$ as $\omega \to 0$, so the integral is logarithmically divergent. Thus, $\lVert K \mc{A}\rVert_{\ms{H}_{\text{R}}^{+}}^{2}= \infty$. Therefore, if Alice displaces a charged particle to a different orbit of the boost Killing field and the particle remains on this new uniformly accelerated trajectory forever, an infinite number of ``soft horizon photons'' will be radiated through the Rindler horizon\footnote{ These ``soft horizon photons'' are closely related to the ``soft hair'' discussed by Hawking, Perry, and Strominger \cite{Hawking:2016msc} in the case of black hole horizons (see also \cite{Hotta:2000gx}). However, while Hawking, Perry, and Strominger considered effects of matter falling into a black hole, our ``soft radiation'' arises from the displacement of matter sourcing a long range field outside of a horizon. Note that in the case of a black hole, the ``soft radiation'' of Alice's experiment increases the entanglement of the black hole with its exterior.} regardless of how quickly or slowly this process is done. 

The above infrared divergence is an exact mathematical analog of the infrared divergences that occur at null infinity in QED for processes with nonzero memory (see e.g., \cite{asymp-quant,Satishchandran_2019,Carney_2017,carney_2018,Asorey:2018erq}). Note that infrared divergences at null infinity arise only in $d=4$ spacetime dimensions. The reason for this is that in $d$ dimensions, radiation falls off at infinity in null directions as $1/r^{d/2 -1}$, whereas Coulomb fields and associated memory effects fall off as $1/r^{d-3}$, so it is only in $d=4$ dimensions that memory effects occur at radiative order \cite{Pate:2017fgt,Satishchandran_2019}. By contrast, radial Coulomb fields will penetrate a Killing horizon in all spacetime dimensions (see \cite{Garfinkle:2020ktg} for the case of a Schwarzschild black hole) and a displacement of a charge will result in a change in the radial Coulomb field in all dimensions. As analyzed above, this will result in radiation through the horizon in all dimensions high enough for the field in question to admit radiation (i.e., $d \geq 3$ for electromagnetism and $d \geq 4$ for gravity). Consequently, the logarithmic divergence in \cref{Nhor} occurs in all spacetime dimensions that admit radiation\footnote{Indeed, there would also be infrared divergences for a particle that sources a massive field, since the Yukawa field of the particle will also penetrate the horizon}.

Now suppose that Alice displaces the particle a $z$-distance $d\ll D = 1/a$ 
from $D$ to $D' = D +d$ as above, but instead of leaving the particle at $D'$ forever, she leaves it there for proper time\footnote{We have normalized the boost Killing field $b^a$ so that Killing time equals proper time on the orbit at $D$ with acceleration $a$. Since we assume $d = D'-D \ll D$, Killing time and proper time are also (nearly) equal on the orbit at $D'$. Thus, $T$ is also the elapsed Killing time that Alice keeps the particle at $D'$.} $T$ and then returns it to $D$. In this case, the transverse components of the vector potential, $\mc{A}_{A}$, return to their initial values at late times, so there is no ``memory effect'' at the horizon. Correspondingly, there are no infrared divergences in the expected number of photons that propagate through $\ms{H}_{\text{R}}^{+}$. Nevertheless, if $T$ is very large then the expected number of photons $\braket{N}$ will be correspondingly large. To see this, we note that if, for convenience, we work in a gauge where $\mc{A}_A = 0$ initially, then during the era at which the particle is at $D'$, $\mc{A}_A$ will be given by the right side of eq.~(\ref{Aivalue}). If we keep the manner in which the particle is moved from $D$ to $D'$ as well as from $D'$ to $D$ fixed but take $T$ to be very large, the asymptotic behavior of the norm 
eq.~(\ref{eq:innerprod0}) will be dominated by the low-frequency contribution from the era of time $T$ that the particle is displaced. The logarithmic divergence at $\omega = 0$ that would occur if the particle remained at $D'$ forever is now effectively  cut off at frequency $\omega \sim 1/V$, where $V$ denotes the affine time duration on the horizon $\ms{H}_{\text{R}}^{+}$ over which the particle remains at $D'$.  We obtain
\begin{equation}
\label{eq:Rindlernum}
   \braket{N}= \lVert K \mc{A}\rVert^{2}_{\ms{H}_{\text{R}}} \sim q^{2}d^{2}a^{2}\ln\bigg(\frac{V}{\textrm{min}[V_{1},V_{2}]}\bigg)
\end{equation}
where $V_{1},V_{2}\ll V$ are the durations of affine time over which the particle is displaced from $D$ to $D^{\prime}$ and from $D^{\prime}$ back to $D$, so that $1/\textrm{min}[V_{1},V_{2}]$ provides an effective high-frequency cutoff.
However, the affine time $V$ on the horizon is related to boost Killing time on the horizon by
\be
V = V_0 \exp(av)
\ee
and the boost Killing time $v$ corresponds to the proper time $T$ in Alice's lab. Thus, we obtain
\begin{equation}
\label{eq:expNRind}
    \braket{N} \sim q^{2}d^{2}a^{3} T \quad \quad \textrm{(Rindler, EM)} \, .
\end{equation}
Therefore, no matter how slowly the particle is displaced, it is forced to radiate a number of ``soft Rindler horizon photons'' through the Rindler horizon that is proportional to the time $T$ that the particle remains on the displaced trajectory. 

\par We are now in a position to fully analyze Alice's experiment. Alice's lab is uniformly accelerating with acceleration $a$ in Minkowski spacetime. She puts her particle of charge $q$ into a superposition of states separated by $z$-distance $d \ll 1/a$ and keeps these components stationary in her lab for a proper time $T$. She then recombines the components and determines their coherence\footnote{The coherence can be determined as described in \cref{decft}.}. By the analysis of sec.~\ref{raddecoh}, the decoherence is given by eq.~(\ref{decrad}). However, for large $T$, the calculation of $\lVert K \left[G^{\rm ret} (j_1 - j_2) \right]\rVert^2$ corresponds precisely to the calculation we have given above of the number of photons radiated through the Rindler horizon when a charge is displaced for a time $T$. Thus, we obtain
\be
\lVert K \left[G^{\rm ret}(j_1 - j_2) \right]\rVert^2 \sim q^{2}d^{2}a^{3} T.
\ee
In other words, for large $T$, Alice's superposition will decohere due to radiation of ``soft Rindler horizon photons,'' as
\begin{equation}
\label{eq:DRH}
\ms{D}=1-\exp(-\Gamma_{\textrm{rad}}T)
\end{equation}
where the ``decoherence rate'' $\Gamma_\text{rad}$,
is given by, 
\begin{equation}
    \label{eq:rindlerRate}
    \Gamma_\text{rad}=q^{2}d^{2}a^{3}.
\end{equation}

Thus, restoring the constants $c$, $\hbar$, and $\epsilon_0$, Alice's particle will decohere within a time
\begin{align}
\label{emTD}
    T_{\textrm{D}} &\sim \frac{\epsilon_{0}\hbar c^{6}}{a^{3}q^{2}d^{2}} \quad \quad \textrm{(Rindler, EM)}\\
        & \sim 10^{33}~\text{years}~ \bigg(\frac{\textrm{g}}{a}\bigg)^{3}\cdot \bigg(\frac{\textrm{e}}{q}\bigg)^{2} \cdot \bigg(\frac{\rm{m}}{d}\bigg)^{2}.
\end{align}
Thus, if Alice's lab uniformly accelerates at one $g$ in flat spacetime and she separates an electron into two components one meter apart, she would not be able to maintain coherence of the electron for more than $10^{33}$ years. 

A similar analysis holds in the gravitational case\footnote{In the gravitational case, additional stress-energy will be needed to keep Alice's particle in uniform acceleration. We will ignore the gravitational effects of this additional stress-energy.} where Alice separates a massive body with mass $m$ across a distance $d$ and maintains this superposition for a time $T$. In the gravitational case, the ``electric part'' of the perturbed Weyl tensor $E_{ab}=C_{acbd}n^{c}n^{d}$ plays an analogous role to the electric field $E_{a}$ in the electromagnetic version of the gedankenexperiment. For a uniformly accelerating point mass, the only non-vanishing component of the electric part of the Weyl tensor on $\ms{H}^{+}_{\text{R}}$ is $E_{UU}=C_{acbd}\ell^{a}n^{c}\ell^{b}n^{d}$. 

Gravitational radiation on the horizon is described by the pullback, $E_{AB}$,  of $E_{ab}$, which vanishes for the static point mass. However, the process of quasistatically moving the static point mass involves a change in $E_{UU}$ on $\ms{H}^{+}_{\text{R}}$. The (once-contracted) Bianchi identity on the horizon yields 
\begin{equation}
\label{eq:Bianchi2}
    \mathcal{D}^{A}E_{AB}=\partial_{V}E_{UB}, \quad \quad 
    \mathcal{D}^{A}E_{UA} = \partial_{V}E_{UU}
\end{equation}
which implies that 
\begin{equation}
\label{eq:bianchi}
     \mathcal{D}^{A} \mathcal{D}^{B}E_{AB} = \partial_{V}^{2}E_{UU}
\end{equation}
which is closely analogous to eq.~(\ref{eq:MaxEq}). As in the electromagnetic case, if a uniformly accelerating point mass is quasistatically moved there is necessarily gravitational radiation through $\ms{H}_{\text{R}}^{+}$. 

To determine the number of ``Rindler horizon gravitons'' emitted we quantize the linearized gravitational field. For a metric perturbation $h_{ab}$ in a gauge where $h_{ab}n^{a}=0$ and $\delta^{AB}h_{AB}=0$, the ``free data'' on $\ms{H}_{\text{R}}^{+}$ is $h_{AB}$. A ``particle'' in the standard Fock space associated to the Poincaré invariant vacuum is then a positive frequency solution with respect to affine parameter $V$ and the inner product on the one-particle Hilbert space is given by a direct analog of eq.~(\ref{eq:innerprod0}) with the vector potential $\mc{A}_{A}$ replaced with the metric perturbation $h_{AB}$, namely
\begin{equation}
\label{eq:innerprodh}
    \braket{K h_{1}| \,K h_{2}}_{\ms{H}_{\text{R}}^{+}}=\frac{1}{8}\int_{\bb{R}^{2}}dxdy \int_{0}^{\infty}\frac{\omega d\omega}{2\pi}\delta^{AB}\delta^{CD}\overline{{\hat{h}_{1,AC}}} \hat{h}_{2,BD}.
\end{equation}
Finally, $E_{AB}$ is related to the metric perturbation $h_{AB}$ by 
\be
E_{AB}=-\frac{1}{2}\partial_{V}^{2}h_{AB} \, .
\label{ehrel}
\ee
Equations (\ref{eq:bianchi}) and (\ref{ehrel}) directly imply that a permanent change, $\Delta E_{UU} \neq 0$, in the $U$-$U$ component of the electric part of the Weyl tensor on $\ms{H}_{\text{R}}^{+}$ implies
a permanent change, $\Delta h_{AB} \neq 0$, in the perturbed metric on $\ms{H}_{\text{R}}^{+}$ between early and late times. In the quantum theory, as in the electromagnetic case, this implies a logarithmic infrared divergence in the number of gravitons emitted through $\ms{H}_{\text{R}}^{+}$ in the process where a uniformly accelerating charge is moved to a new orbit of the same boost Killing field and then remains at the new position forever. 

The analysis of Alice's experiment proceeds in a similar manner to the electromagnetic case. Alice does not maintain the relative separation of her wavefunction forever but closes her superposition after a proper time $T$. As before, the number of entangling gravitons emitted to the Rindler horizon is logarithmically growing in affine time and therefore linearly growing in the proper time duration $T$ of Alice's experiment. We obtain 
\begin{equation}
\braket{N} \sim m^{2}d^{4}a^{5}T \quad \quad \textrm{(Rindler, GR)} \, .
\end{equation}  
Thus, restoring constants, we find that the Rindler horizon decoheres the quantum superposition of a uniformly accelerating massive body in a time 
\begin{align}
\label{eq:DecGRacc}
 T_{D}^{\textrm{GR}}\sim& \frac{\hbar c^{10}}{Gm^{2}d^{4}a^{5}}  \quad \quad \textrm{(Rindler, GR)}  \\ 
 \sim &~2~\textrm{fs}~\bigg(\frac{\textrm{M}_{\textrm{Moon}}}{m}\bigg)^{2}\cdot \bigg(\frac{\textrm{R}_{\textrm{Moon}}}{d}\bigg)^{4} \cdot \bigg(\frac{\rm{g}}{a}\bigg)^{5}.
\end{align}
Therefore, if the Moon were accelerating at one $g$ and occupied a quantum state with its center of mass superposed by a spatial separation of the order of its own radius then it would decohere within about 2 femtoseconds. Of course, it would not be easy to put the moon in such a coherent quantum superposition.

Note the acceleration of a stationary observer outside of a black hole who is reasonably far\footnote{It should be emphasized that the estimates made in \cite{Danielson_2022} that yielded eqs.(\ref{eq:NEBh}) and (\ref{NGRbh}) assumed that Alice's lab is reasonably far from the black hole. If Alice's lab is extremely close to the black hole (i.e., at a distance $D \ll M$ from the horizon), then the black hole analysis would reduce to the Rindler case analyzed here.} ($D \gtrsim M$) from the event horizon is $a\sim M/D^{2}$. If we substitute $a =  M/D^{2}$ in eqs. (\ref{emTD}) and (\ref{eq:DecGRacc}), we obtain eqs.~(\ref{eq:NEBh}) and (\ref{NGRbh}), respectively. Therefore, it might be tempting to believe that what is important in all cases is the acceleration of Alice's lab. However, this is not the case. In particular, if we replace the black hole by an ordinary star (and if there are no dissipative effects in the star), then there will not be any analogous decoherence effect, even though the acceleration of Alice's lab is the same as in the case of a black hole. Furthermore, as we shall see in 
sec.~\ref{cosdecoh}, decoherence effects associated with the cosmological horizon occur in de Sitter spacetime even for nonaccelerating observers. It is the presence of a Killing horizon that is the essential ingredient for the fundamental rate of decoherence of quantum superpositions as described in this paper.

We now consider another potential cause of decoherence, namely Unruh radiation.

\subsection{Decoherence Due to Scattering of Unruh Radiation}
\label{subsec:thermalRate}

The Minkowski vacuum state restricted to a Rindler wedge is a thermal state at the Unruh temperature
\be
\mc{T} = \frac{ a}{2 \pi }
\label{untem}
\ee
relative to the notion of time translations defined by the Lorentz boost Killing field $b^a$, eq.~(\ref{boostkvf}). Thus, the superposition state of Alice's particle will be buffeted by this thermal bath of Unruh radiation. Scattering of this radiation will cause some decoherence of Alice's particle. Indeed, since this decoherence should occur at a steady rate while the superposition is kept stationary (and thus the decoherence will be proportional to $T$), one might even suspect that scattering of Unruh radiation could be the same effect as found in the previous section but expressed in a different language. The purpose of this subsection is to show that this is not the case, i.e., decoherence due to scattering of Unruh radiation and decoherence due to radiation of ``soft'' photons/gravitons through the horizon are distinct effects. Furthermore, we shall show that, for reasonable parameter choices, the decoherence rate due to the scattering of Unruh radiation is smaller than the decoherence rate due to emitted radiation as obtained in the previous section. We will consider only the electromagnetic case in this subsection.

The decoherence rate of a spatial superposition due to collisions with particles in an environment has been analyzed in \cite{collisional_Joos, collisional_Gallis, collisional_Diosi, collisional_Hornberger}, and we will adapt this analysis to obtain a rough estimate of the decoherence caused by the scattering of Unruh radiation. As in eq.~(\ref{sup12}), Alice has a particle of charge $q$ in a state $\ket{\psi} = (\ket{\psi_1} + \ket{\psi_2})/\sqrt{2}$, where $\ket{\psi_1}$ and $\ket{\psi_2}$ are spatially separated by a distance $d$. Since we require $d \ll 1/a$ (see eq.~(\ref{dllD})) and since the typical wavelength of Unruh photons at temperature eq.~(\ref{untem}) is $\lambda \sim 1/a$, we are in the scattering regime where $\lambda \gg d$. In an elastic scattering event between Alice's particle and a photon in the Unruh radiation, the final outgoing state of the photon will depend upon which branch of the superposition the photon scattered off of. Let $\ket{\chi_1}$ denote the outgoing state of the Unruh photon for scattering off of $\ket{\psi_1}$ and let  $\ket{\chi_2}$ denote the outgoing state for scattering off of $\ket{\psi_2}$. Decoherence will occur to the extent to which these outgoing states of the scattered Unruh photon are distinguishable, i.e., ${\mathscr D} =1 - |\braket{\chi_1|\chi_2}|$. 

In order to obtain a rough estimate of the decoherence resulting from a single scattering event, we consider the corresponding Minkowski process of the scattering of a photon of momentum $p$ off of an inertial superposition separated by $d$, with $d \ll 1/p$. Assuming that the charged particle states $\ket{\psi_1}$ and $\ket{\psi_2}$ are identical except for their location, the scattered photon states $\ket{\chi_1}$ and $\ket{\chi_2}$ should differ only by the action of the translation operator $e^{-i \vec{\mc{P}} \cdot \vec{d}}$, i.e.,
\be
\ket{\chi_2} \approx e^{-i \vec{\mc{P}} \cdot \vec{d}} \ket{\chi_1}
\ee
where $\vec{\mc {P}}$ denotes the photon momentum operator. 
Expanding the exponential, we obtain the following rough estimate of the decoherence resulting from a single scattering event involving a photon of momentum $p$
\be
1  - |\braket{\chi_1|\chi_2}| \sim p^2 d^2
\label{scdec}
\ee
where we have ignored any dependence on the angle between the incoming momentum $\vec{p}$ and the separation $\vec{d}$.
We will take eq.~(\ref{scdec}) as our estimate of the decoherence of Alice's particle resulting from the scattering of a single Unruh photon of ``Rindler momentum'' $p$ (i.e., of energy $\epsilon = p$ with respect to the boost Killing field $b^a$). 

The total decoherence rate due to scattering of Unruh radiation is then given by
\begin{equation}
    \Gamma_\text{scatt} \sim d^2\int_0^\infty dp~ p^2\varrho(p)  \sigma(p)    
\end{equation}
where $\varrho(p)$ is the number density of photons at momentum $p$ (so $\varrho(p)$ is also the incoming flux of photons) and $\sigma(p)$ is the scattering cross-section. 
For a thermal distribution of photons\footnote{The factor of $p^2$ in the numerator of eq.~(\ref{planck}) arises from the density of states in Minkowski spacetime. We ignore here any differences between the Minkowski and Rindler densities of states.} we have
\begin{equation}
    \varrho(p) \sim \frac{p^2}{e^{p/\mc{T}}-1}.
    \label{planck}
\end{equation}
We take $\sigma$ to be given by the Thomson cross-section
\begin{equation}
    \sigma = \frac{8\pi}{3}\frac{q^4}{(4\pi m )^2},
\end{equation} 
where $m$ is the mass of Alice's particle. Putting this all together, our estimate of the decoherence rate due to scattering of Unruh photons is
\begin{equation}
   \Gamma_\text{scatt} \sim  \frac{q^4d^2 a^5}{m^2} \quad \quad \textrm{(Rindler, EM)} \, .
   \label{gamscat}
\end{equation}

Comparing eq.~(\ref{gamscat}) to the rate of decoherence, $ \Gamma_\text{rad}$ due to the emission of soft photons given by eq.~(\ref{eq:rindlerRate}), one can immediately see that the effects are distinct. In particular, $ \Gamma_\text{rad}$ has no dependence on the mass, $m$, of Alice's particle, whereas $\Gamma_\text{scatt}$ does depend on $m$ on account of the mass dependence of the scattering cross-section. The ratio of these decoherence rates is given by
\be
\frac{ \Gamma_\text{scatt}}{ \Gamma_\text{rad}} \sim \frac{q^2 a^2}{m^2} = \left( \frac{q/m}{D} \right)^2
\ee
Now, $q/m$ is the ``charge radius'' of Alice's particle and, as argued in \cite{Belenchia_2018}, it represents a fundamental lower bound to the spread of a charged particle due to vacuum fluctuations of the electromagnetic field. Therefore, in order that $\ket{\psi_1}$ and $\ket{\psi_2}$ not overlap, we must have $d > q/m$. Since $d \ll D$, we conclude that 
\be
\frac{ \Gamma_\text{scatt}}{ \Gamma_\text{rad}} \ll 1
\ee
i.e., the contribution to decoherence from the scattering of Unruh radiation is negligible compared with the decoherence due to emission of soft photons through the Rindler horizon. 

A similar analysis holds for a charged particle superposition outside of a black hole. It is worth noting, that the decoherence effects due to scattering of Hawking radiation will decrease with distance, $D$, from the black hole only as $1/D^2$ for large $D$, giving,
\begin{equation}
\Gamma_\mathrm{scatt}\sim
\frac{q^4d^2}{m^2M^3}\frac{1}{D^2}
\quad \textrm{(black hole, EM)}.
\end{equation}
On the other hand, by eq.~(\ref{eq:NEBh}) the decoherence effects of radiation of soft photons through the horizon decreases with $D$ as $1/D^6$. Thus at sufficiently large $D$, the decoherence effects due to scattering of Hawking radiation will dominate. However, in this regime, both effects are extremely small.

\subsection{ Decoherence From the Inertial Perspective}
\label{subsec:decinertialperp}

In our analysis of the decoherence of a spatial superposition in the presence of a black hole \cite{Danielson_2022} as well as in our analysis of the decoherence of a spatial superposition in Rindler spacetime given above in sec.~\ref{subsec:softrad}, it may appear that we have introduced a radical new mechanism for decoherence, namely radiation of soft photons and gravitons through a horizon. The main purpose of this subsection is to show that, in fact, the decoherence we derived in the Rindler case can also be obtained by entirely conventional means. In the Rindler case, we are simply considering a uniformly accelerating superposition in Minkowski spacetime. The radiation of entangling photons to infinity from such a superposition can be calculated in the inertial viewpoint by standard methods, without introducing concepts such as a Rindler horizon.
It is instructive to calculate the decoherence from the inertial viewpoint both in order to validate the results of sec.~\ref{subsec:softrad} as well as to gain insight into how the emitted ``soft photons'' would be interpreted by an inertial observer. As we shall see, the entangling photons as seen by a faraway inertial observer along the forward axis of acceleration will be ``hard'' even though, from her point of view, Alice has performed the experiment adiabatically. We will restrict our analysis in this subsection to the electromagnetic case. 

The Li\'enard-Wiechert solution for the potential of a point charge in Minkowski spacetime following an arbitrary worldline $X^\mu(\tau)$ is, in Lorenz gauge,
\begin{equation}
    A^{\mu}(x)=\frac{1}{4\pi}\frac{1}{\alpha}\frac{q}{|\vec{x}-\vec{X}(t_{\text{ret}})|}\frac{dX^{\mu}}{dt}(t_{\text{ret}})
    \label{lwsol}
\end{equation}
where 
\begin{equation}
    \alpha \equiv 1 - \hat{n}\cdot \frac{d\vec{X}}{dt}(t_{\text{ret}}) \quad \text{and } \hat{n}=\frac{\vec{x}-\vec{X}(t_{\text{ret}})}{|\vec{x}-\vec{X}(t_{\text{ret}})|}.
\end{equation}
For a uniformly accelerated trajectory with acceleration $a$, we have
\begin{equation}
    X^{\mu}(\tau)=  \bigg(\frac{1}{a}\sinh(a\tau),0,0,\frac{1}{a}\cosh(a\tau)\bigg).
    \label{accwl}
\end{equation}
In Bondi coordinates $(u,r,\theta,\phi)$ with
\be
u \equiv t - r
\ee
the future light cone of an event at proper time $\tau$ on the worldline eq.~(\ref{accwl}) reaches null infinity at
\begin{equation}
\label{eq:retardedTime}
    a u=\sinh(a\tau)-\cos\theta\cosh(a\tau).
\end{equation}

Electromagnetic radiation is described by the pullback of the electromagnetic field, eq.~(\ref{lwsol}), to null infinity. Taking the limit as $r \to \infty$ at fixed $u$, we obtain\footnote{The vector potential is not smooth at $\scri^{+}$ in Lorenz gauge but one can do an asymptotic gauge transformation such that $A_{a}$ is smooth at $\scri^{+}$. Such a gauge transformation does not affect the angular components $A_{A}$ at $\scri^{+}$ \cite{Satishchandran_2019}, so we can calculate $A_A$ using our Lorenz gauge expression.}
\begin{equation}
    A_{A}(u,\theta,\phi) =  \frac{-q}{4\pi}\frac{\sinh(a\tau)\sin\theta}{\cosh(a \tau)-\cos\theta\sinh(a\tau)} (d\theta)_{A}
    \label{lwradf}
\end{equation}
where, in this subsection, capital indices from the early alphabet denote angular components on the $2$-sphere cross-sections of $\scri^{+}$. We will be concerned with the difference, at fixed $(u,\theta,\phi)$, between the electromagnetic radiation of a particle following the trajectory eq.~(\ref{accwl}) and a particle following a similar trajectory that is displaced in the $z$-direction by a proper distance $d \ll 1/a$ and thus has
\be
\delta a= a^2 d.
\ee
We denote this difference by 
\be
A^{\text{d}}_A(u,\theta,\phi) \equiv {A_A (a+\delta a) } -{A_A (a)} \approx \delta a \left( \frac{\partial A_A}{\partial a} \right)_{u, \theta}  
\ee
From eq.~(\ref{lwradf}), we obtain
\be
A^{\text{d}}_A = -\frac{a^2qd}{4\pi}\frac{u\sin\theta}{(\cosh(a\tau)-\cos\theta \sinh(a\tau))^{3}}(d\theta)_{A}
     \label{eq:dipole}
\ee
where eq.~(\ref{eq:retardedTime}) was used to compute $(\partial \tau/\partial a)_{(u, \theta)}$.

In her experiment, Alice starts with her particle in a uniformly accelerating state. Over a proper time $T_1$, she separates it into two uniformly accelerating components separated by a distance $d$ as above. She keeps these components separated for a proper time $T$, and she then recombines them over a proper time $T_2$. The difference between the radiation fields of these components is given by
\be
\mc{A}_A \equiv \mc{A}_{1,A} - \mc{A}_{2,A} = F(\tau) A^{\text{d}}_A
\ee
where the smooth function $F$ is such that $F(\tau) = 0$ for $\tau < - T_1$ and $\tau > T + T_2$, whereas $F(\tau) = 1$ for $0 < \tau < T$. The entangling photon content is then given by
\be
\braket{N} = \lVert K \mc{A}\rVert^2 = 2 \int_{\mathbb{S}^2} d\Omega \int_0^\infty \frac{\omega d\omega}{2\pi}   ~ \overline{\hat{\mc{A}}_A}\hat{\mc{A}}^A
\label{kgnorm}
\ee
where $\hat{\mc{A}}_A(\omega,\theta,\phi)$ denotes the Fourier transform of $\mc{A}_A(u,\theta,\phi)$ with respect to $u$, i.e.,
\be
\hat{\mc{A}}_A(\omega,\theta,\phi) = \int_{-\infty}^{\infty} du~e^{i \omega u} \mc{A}_A(u,\theta,\phi).
\ee
We are interested in estimating $\braket{N}$ for large $T$.

In order to evaluate the Fourier transform integral, it is useful to note that, at fixed $a$, we have
\be
\frac{du}{d\tau} = \cosh(a \tau) - \cos \theta \sinh (a \tau)
\ee
and
\be
\frac{d^2u}{d\tau^2} = a^2 u.
\ee
It follows that 
\begin{align}
\frac{d}{du} \left( \frac{1}{du/d\tau} \right) &=  \frac{1}{du/d\tau} \frac{d}{d\tau} \left( \frac{1}{du/d\tau} \right) \nonumber \\
&= \frac{-a^2 u}{\left( \cosh(a \tau) - \cos \theta \sinh (a \tau) \right)^3}
\end{align}
Thus, we have
\be
A^{\text{d}}_A =  \frac{qd \sin\theta}{4 \pi} (d\theta)_A \frac{d}{du} \left( \frac{1}{du/d\tau} \right) 
\ee
and
\be
\hat{\mc{A}}_A = \frac{qd\sin\theta }{4 \pi } (d\theta)_A \int_{-\infty}^{\infty} du~ e^{i \omega u} F(\tau) \frac{d}{du} \left( \frac{1}{du/d\tau} \right).
\ee
Integrating by parts, we obtain
\begin{align}
    \hat{\mc{A}}_A(\omega,x^{A})=& - \frac{qd\sin\theta}{4\pi}(d\theta)_{A}\bigg[i\omega \int_{-\infty}^{\infty}du~e^{i\omega u}\frac{F(\tau)}{du/d\tau}\nonumber \\
    &+ \int_{-\infty}^{\infty}du~e^{i\omega u}\frac{F^{\prime}(\tau)}{(du/d\tau)^{2} }\bigg].
    \label{eq:fourier}
\end{align}
The second term in this equation contributes only during the time intervals $(-T_1, 0)$ and $(T, T + T_2)$ when Alice opens and closes the superposition. For large $T$, its contribution can be shown to be negligible compared with the first term. Therefore, we have
\be
\hat{\mc{A}}_A(\omega,x^{A}) \approx -(d\theta)_{A} \frac{i \omega qd\sin\theta}{4\pi} I
\label{ai}
\ee
where
\be
I \equiv \int_{-\infty}^{\infty}du~e^{i\omega u}\frac{F(\tau)}{du/d\tau}.
\ee

To evaluate $I$, we approximate $F$ by a step function in the $\tau$-interval $[0,T]$. The corresponding interval, $[u_0, u_T]$, in $u$ is
\begin{align}
    u_0 &= -\frac{1}{a} \cos \theta \nonumber \\
    u_T &= \frac{1}{2a} \left[ e^{aT}(1-\cos \theta)-e^{-aT}(1+\cos\theta) \right]. 
    \label{intlim}
\end{align}
Noting that \begin{equation}
   \frac{du}{d\tau} = \sqrt{a^{2}u^{2}+\sin^{2}\theta}
    \end{equation}
we obtain
\be
 I  \approx \int_{u_0}^{u_T}du~\frac{e^{i\omega u}}{\sqrt{a^2u^2+\sin^2\theta}}.
 \ee
It can be seen that for large $T$, the dominant contribution to $I$ will come from small angles, $\theta \ll 1$. For $aT \gg 1$, the upper limit of the integral may then be approximated as
\begin{eqnarray}
   u_T  & \approx& \frac{1}{4a} e^{aT}\theta^2- \frac{1}{a}e^{-aT} \quad \textrm{for} \,\, \theta\ll 1  \nonumber \\
    &\sim& \begin{cases} 
      0 &\text{for } \theta^2/4 < e^{-aT} \\
       \frac{1}{4a} \theta^2e^{aT}&\text{for }  \theta^2/4 \ge e^{-aT} 
   \end{cases}.
\end{eqnarray}
For $aT \gg 1$, the contribution to $I$ from $\theta^2/4 < e^{-aT}$ can be shown to make a negligible contribution to $\braket{N}$, eq.~(\ref{kgnorm}). Therefore, we may approximate $I$ as 
\be
I \sim \Theta(\theta^2-4e^{-aT})\int_{-1/a}^{\exp(aT)\theta^2/(4a)} du~ \frac{e^{i\omega u}}{\sqrt{a^2u^2+\sin^2\theta}}
\label{int3}
\ee
where 
\be
\Theta (x) \equiv \begin{cases}
0 & \text{for } x < 0 \\
       1 &\text{for }  x \ge 0.
   \end{cases}
\ee
For $0 <\omega < 4ae^{-aT}/\theta^2$, we may bound $I$ by replacing $e^{i\omega u}$ by $1$. The integral can then be evaluated explicitly, and it can be shown that for $aT \gg 1$, the contribution to $\braket{N}$ from this frequency range is negligible. For $\omega > 4ae^{-aT}/\theta^2$, the integrand is oscillatory for $u > \exp(aT)\theta^2/(4a)$, and, for $aT \gg 1$, we will make negligible error in our estimate of $\braket{N}$ if we replace the upper limit of eq.~(\ref{int3}) by $\infty$. We will also make a negligible error by replacing the lower limit by $0$. Thus, for $aT \gg 1$, we may approximate $I$ as
\begin{equation}
  I   \sim \Theta(\theta^2-4e^{-aT})\Theta(\omega-4ae^{-aT}/\theta^2)\int_0^{\infty} du ~\frac{e^{i\omega u}}{\sqrt{a^2u^2+\sin^2\theta}}.
\end{equation}
Evaluating the integral we obtain
\begin{align}
  I \sim   \frac{1}{a}  \Theta(\theta^2 & -4e^{-aT})\Theta(\omega - 4ae^{-aT}/\theta^2)\bigg(\frac{1}{2} i \pi  I_0(  \sin \theta \omega/a ) \nonumber
         \\+&K_0(  \sin \theta\omega/a )-\frac{1}{2} i \pi  \pmb{L}_0(  \sin \theta\omega/a )\bigg)
\end{align}
where $I_0, K_0$ are Bessel functions and $\pmb{L}_0$ is a Struve function. This expression is highly suppressed for $\omega > a/\theta$, so we can expand in $ \theta \omega/a$ and truncate the function above $\omega = a/\theta$ to obtain,
\begin{equation}
 I \sim-\frac{1}{a}  \Theta(1-\theta\omega/a)\Theta(\theta^2-4e^{-aT})\Theta(\omega-4ae^{-aT}/\theta^2) \ln \left(\theta\omega/a\right).
\label{eq:fourier2}
\end{equation}
Note that the restrictions $\omega<a/\theta$, and $\theta>2e^{-aT/2}$ imply a frequency cutoff at $\omega \sim a e^{aT/2}/2$. By eqs.(\ref{eq:fourier2}) and (\ref{ai}), the frequency spectrum of $\hat{\mc{A}}_A$ goes as $\omega\ln(\omega/a)$ up to this cutoff, i.e., the spectrum is ``hard'' and becomes increasingly so for large $T$. This contrasts with the increasingly ``soft'' spectrum on the Rindler horizon, which goes as $1/\omega$ down to a low frequency cutoff $\sim 1/V \propto e^{-aT}$. Thus, the ``soft horizon photons'' from the Rindler perspective are ``hard'' photons from the inertial perspective.

From eq.~(\ref{kgnorm}) for $\braket{N}$ together with our expression eq.~(\ref{ai}) for $\hat{\mc{A}}_A$ and the expression 
eq.~(\ref{eq:fourier2}) that we have just derived for $I$, we obtain
\begin{equation}
\braket{N} \sim \left(\frac{qd}{a} \right)^2 \int d\omega d\theta~ \theta^3 \omega^3  \left(\ln \frac{\omega \theta}{a}\right)^2
\label{N1}
\end{equation}
where the region of $\omega$-$\theta$ integration is determined by the $\Theta$-functions appearing in eq.~(\ref{eq:fourier2}) as well as the geometrical restriction $\theta \lesssim 1$. We can break up this region into the portion with $\omega \leq a$ and the portion with $\omega > a$. Since the region with $\omega \leq a$ and $\theta \lesssim 1$ is bounded and the integrand of eq.~(\ref{N1}) is bounded in this region, the contribution to $\braket{N}$ from $\omega \lesssim a$ is bounded by a constant that is independent of $T$. We may therefore discard this contribution. In the region $\omega > a$, the third $\Theta$-function in eq.~(\ref{eq:fourier2}) is redundant, and the integration region is
\begin{eqnarray}
a \leq &\omega& \leq a e^{aT/2}/2 \\
2 e^{-aT/2} \leq &\theta& \leq \frac{a}{\omega}.
\end{eqnarray}
For $a T \gg 1$, we will make negligible error by replacing the lower limit of $\theta$ by $0$. We thereby obtain
\be
\braket{N} \sim \left(\frac{qd}{a} \right)^2  \int_a^{a \exp(aT/2)/2} d\omega  \int_0^{a/\omega}d\theta ~\theta^3 \omega^3  \left(\ln \frac{\omega \theta}{a}\right)^2.
\ee
Making the change of variables from $\theta$ to 
\be
x = \frac{\omega}{a} \theta
\ee
we find that the $\theta$-integral becomes
\be
 \int_0^{a/\omega}d\theta~\theta^3 \omega^3  \left(\ln \frac{\omega \theta}{a}\right)^2 = \frac{a}{\omega} a^3 \int_0^1 dx~ x^3 (\ln x)^2 \sim \frac{a^4}{\omega}.
\ee
Thus, we obtain
\begin{eqnarray}
\braket{N} 
& \sim&  \left(\frac{qd}{a} \right)^2 a^4 \int_a^{a\exp(aT/2)/2}  \frac{d\omega}{\omega}   \nonumber \\
& \sim  & a^2 q^2 d^2 \ln [\exp(aT/2)] \nonumber \\
&\sim& a^3 q^2 d^2 T.
\end{eqnarray}
This estimate agrees with eq.~(\ref{eq:expNRind}). 

Thus, we have succeeded---with considerable effort!---in our goal of deriving the decoherence of Alice's superposition by entirely conventional means. It is notable how much simpler the calculation of sec.~\ref{subsec:softrad} was compared to the calculation that we have just completed.

\section{Cosmological Horizons Decohere Quantum Superpositions}
\label{cosdecoh}

In this section, we apply our analysis to de Sitter spacetime. The de Sitter metric in a static patch is given by
\begin{equation}
ds^2 = - f(r) dt^{2} + f(r)^{-1}dr^{2} + r^{2}q_{AB}dx^{A}dx^{B}
\end{equation}
where, in this section, $x^{A}$ are angular coordinates on the $2$-sphere, $q_{AB}$ is the unit round metric on the $2$-sphere, and
\be
f(r) = 1 - r^{2}/R_H^{2}
\ee
where $R_H$ (the ``Hubble radius'') is a constant. The coordinate singularity at $r = R_H$ corresponds to the ``cosmological horizon,'' which is a Killing horizon of the static Killing field $(\partial/\partial t)^{a}$. The relation between ``affine time,'' $V$, and ``Killing time,'' $v$, on the future cosmological horizon is 
\begin{equation}
V = e^{v/R_H}.
\end{equation}

The general analysis of sec.~\ref{raddecoh} applies to the decoherence of a static superposition in de Sitter spacetime. The estimates of the decoherence due to emission of soft photons and gravitons through the cosmological horizon when Alice keeps the superposition present for a time $T$ can be made in exact parallel with the analysis of sec.~\ref{subsec:decacc} in the Rindler case and \cite{Danielson_2022} in the black hole case. The only noteworthy new ingredient in de Sitter spacetime is that the worldline $r=0$ is an orbit of the static Killing field that is inertial, i.e., non-accelerating. We now estimate the decoherence of a spatial superposition created in Alice's lab at $r=0$ and thereby show that decoherence will occur even though Alice's lab is not accelerating.

By Gauss' law, a point charge placed at $r=0$ will give rise to a radial electric field $E_U$ on the future cosmological horizon given by
\be
E_U \sim \frac{q}{R_H^2}
\ee
where $E_U = F_{ab} \ell^{a}n^{b}$ on the horizon with $n^{a}=(\partial/\partial V)^a$ tangent to the affinely parametrized null generators of the horizon and $\ell^{a}= (\partial/\partial U)^a$ a radial null vector with $n^{a}\ell_{a} = -1$.
The change in the electric field on the horizon resulting from a displacement of the charge to $r=d \ll R_H$ is
\be
\Delta E_U \sim \frac{q d}{ R_H^3}.
\ee
By paralleling the steps that led to eq.~(\ref{Aivalue}) above, we find that the change in the tangential components of the vector potential at the horizon is
\begin{equation}
|\Delta \mc{A}_{A}| \equiv \left(  R^{-2}_H q^{AB} \Delta \mc{A}_{A} \Delta \mc{A}_{B} \right)^{1/2} \sim \frac{qd}{R_H^{2}}.
\end{equation}
By paralleling the steps that led to eq.~(\ref{eq:expNRind})---assuming that the electromagnetic field is initially in the de Sitter invariant vacuum (see \cref{dsinv})---we obtain the estimate
\begin{equation}
\braket{N} \sim \frac{q^{2}d^{2}}{ R_H^3}T \quad \quad \textrm{(de Sitter, EM)} \, .
\end{equation}
Thus, restoring constants, the decoherence time due to the presence of the cosmological horizon is 
\be
T_{\text{D}} \sim \frac{\hbar \epsilon_{0} R_H^3}{q^{2}d^{2}} \quad \quad \textrm{(de Sitter, EM)} \, .
\ee
Since $d \ll R_H$, the decoherence time will be much larger than the Hubble time $R_H/c$ unless $q$ is extremely large relative to the Planck charge $q_P \equiv \sqrt{\epsilon_0 \hbar c}$. Nevertheless, we see that decoherence does occur despite the fact that Alice's lab is inertial.
\par A similar analysis applies in the gravitational case for a spatial superposition of a massive particle in Alice's lab at $r=0$. In parallel with the derivation given in sec.~\ref{subsec:softrad} above, we find
\begin{equation}
\braket{N} \sim \frac{m^{2}d^{4}}{ R_H^{5}}T  \quad \quad \textrm{(de Sitter, GR)}
\end{equation}
which leads to a decoherence time
\be
  \quad \quad \quad T_{\text{D}}^{\text{GR}}\sim \frac{\hbar R_H^{5}}{G m^{2}d^{4}}  \quad \quad \textrm{(de Sitter, GR)} \, .
\ee
\vspace{5px}

\begin{acknowledgments}
D.L.D. acknowledges support as a Fannie and John Hertz Foundation Fellow holding the Barbara Ann Canavan Fellowship and as an Eckhardt Graduate Scholar at the University of Chicago. This research was supported in part by NSF Grant No. 21-05878 to the University of Chicago and the Princeton Gravity Initiative at Princeton University.
\end{acknowledgments}
\bibliography{Cosm_dec}
\end{document}